\documentclass[preprint,prb]{revtex4}
\usepackage{graphicx}
\begin{document}

\title{Comparative analysis of different
 preparation methods of chalcogenide glasses:
 Molecular dynamics structure simulations}
\author{J. Heged\"us$^1$,  K. Kohary$^{1,2}$, S. Kugler$^1$} 
\affiliation{$^1$Department of Theoretical Physics,
Budapest University of Technology and Economics,
H-1521 Budapest, Hungary,}
\affiliation{$^2$Department of  Materials, University of Oxford,
Parks Road, Oxford, OX1 3PH, UK}


\begin{abstract} 

Two different preparation methods (liquid-quenching and
evaporation) of  chalcogenide glasses have been investigated
by  molecular dynamics simulations. Our particular aim was
to determine how the structural changes occur due to the different 
preparation methods. We applied a 
 classical  empirical three-body potential of selenium 
to describe the interactions between atoms.
 Our simulation shows that a significant difference
can be observed in the homogeneities.


\end{abstract}

\maketitle
\section{Introduction}

Chalcogenide glasses  have been  the subject of  numerous experimental
works in the recent decades. Basically, there are two different ways 
to   produce    samples   for   experiments;    liquid-quenching   and
evaporation. In the  first case the initial phase  of raw materials is
liquid while  in the latter  case the starting compound  is vaporized.
Usually, the  quenched materials are  named glasses and  the amorphous
forms  are prepared  from  gas phase  onto  substrates. The  principal
advantage  of rapid  quenching compared to evaporation  is  that the
method can provide large volume  of samples.  There may be differences
in  the physical  properties  of samples  produced  by different  ways
because these states are non-equilibrium states \cite{Morigaki99}.  Our
particular aim was  to determine how the structural  changes occur due
to the different  preparation methods.  In order to  obtain an answer for
this question  we performed molecular dynamics simulations.
Our atomic  networks contained  about 1000 selenium  atoms interacting
via  classical empirical  three-body  potential \cite{Oligschleger96}.
Non-crystalline selenium has received  particular attention since it is
the  model material of  twofold  coordinated covalently  bounded  chalcogenide
glasses.

\section{Simulation details}

There are  two main possibilities  for structural modeling  on the atomic
scale.  First is Monte Carlo  (MC) type methods.  Traditional MC using
a  potential  minimizes  the  total  energy  in  energy  hyper-surface.
Recently a  new version of this  method - the  so-called Reverse Monte
Carlo (RMC) simulation  - has been developed which  is also convenient
for investigation of  amorphous materials.  It is based  on results of
diffraction  measurements.  This method  was applied  for constructing
large  scale  a-Si and  a-Se models  \cite{kugler93,jovari03}.
Second,  Molecular  Dynamics (MD)  also  needs  a  local potential  to
describe  the  interaction between  atoms.   We  have  developed a  MD
computer  code (ATOMDEP program  package) to  simulate real  preparation
 procedure  of disordered structures.

\subsection{Computer simulation of preparations}

 Amorphous and glassy structures are usually
grown by different vapor depositions  on substrates.  In our recent MD
work \cite{Kohary01},  the growth of amorphous carbon  films was simulated
by this  method. Only a brief  summary of our  simulation technique is
given here  (for details,  see Ref.  \cite{Kohary01}).   A crystalline
lattice cell containing  324 selenium atoms was employed  to mimic the
substrate.  There were 108 fixed atoms at the bottom of the substrate.
The  remaining atoms could  move with  full dynamics.   The simulation
cell was open  along the positive $z$ direction  and periodic boundary
conditions were applied in $x$, $y$ directions.  Kinetic energy of
the atoms in the substrate was rescaled at every MD step~(${\Delta}t$
= 1~fs)  in order to keep  the substrate at  a constant temperature.
In this  kind of simulation there is  no {\it ad hoc}  model for energy
dissipation of incoming  atoms.  In the deposition  process the frequency
of the  atomic injection was  300~fs$^{-1}$.  This  flux is
orders of  magnitude larger than the deposition  rate commonly applied
in the experiments  but we compensate  this disadvantage by  a low substrate
temperature.  After  bombarding (no more incoming atoms)  there were 
30~ps  periods for  structure relaxations in each case.
 Three  different structures
[SeStr]  have been  constructed  by the  technique  mentioned above  at the
temperature of 100~K.  The  average  bombarding energies  of  SeStr0.1,
SeStr1, and SeStr10 models  were 0.1~eV, 1~eV and 10~eV, respectively.

Rapid cooling of   liquid phase is frequently  applied to construct
glassy  structures.   The  system  is  usually  cooled  down  to  room
temperature  by  a  rate  of 10$^{11}$--10$^{16}$~K/s in  computer 
simulations although this rate is some orders 
of magnitude smaller in the experimental techniques.   In  order  to
retrieve  information  on   the  rapid  cooling  (melt-quenching),  we
prepared a model (SeStrQ)
 in the following way. Temperature of a deposited film (SeStr1) 
 was increased up  to 900~K  as an  initial state  (liquid phase),
while the substrate temperature remained the same. After this melting,
the trajectories of the selenium  atoms were followed by full dynamics
for 100~ps.  The  substrate temperature kept at 100~K  leads to the
cooling  of  the film  above  the  substrate.  This technique  can  be
considered as  the computer simulation  of real splat cooling,  where small
droplets of melt are brought into contact with the chill-block.

\subsection{The applied potential}

  Pair potentials can not be  used for covalently bonded structure because
 these types of potentials can not  handle the bond angles. We need at
 least  three body  interactions.   For our  simulation  we applied  a
 classical  empirical three-body potential  \cite{Oligschleger96}. The
 parametrization  of  this potential  is based  on  fitting  the
 structures of  small Se clusters determined by  DFT calculations and
 experimental data due to crystalline phase.

\section{Structural properties}

\subsection{Pair correlation and bond angle distribution functions}

\begin{figure}
\caption{
 Pair
 correlation  functions of our  SeStr1 model  and of a model
constructed by unconstrained RMC
 simulation \cite{jovari03} (experimental)  are shown in  the
 interval of 1-5~\AA. 
}
\vskip 3truecm
\begin{center}
\rotatebox{-90}{
\includegraphics[width=12cm,keepaspectratio]{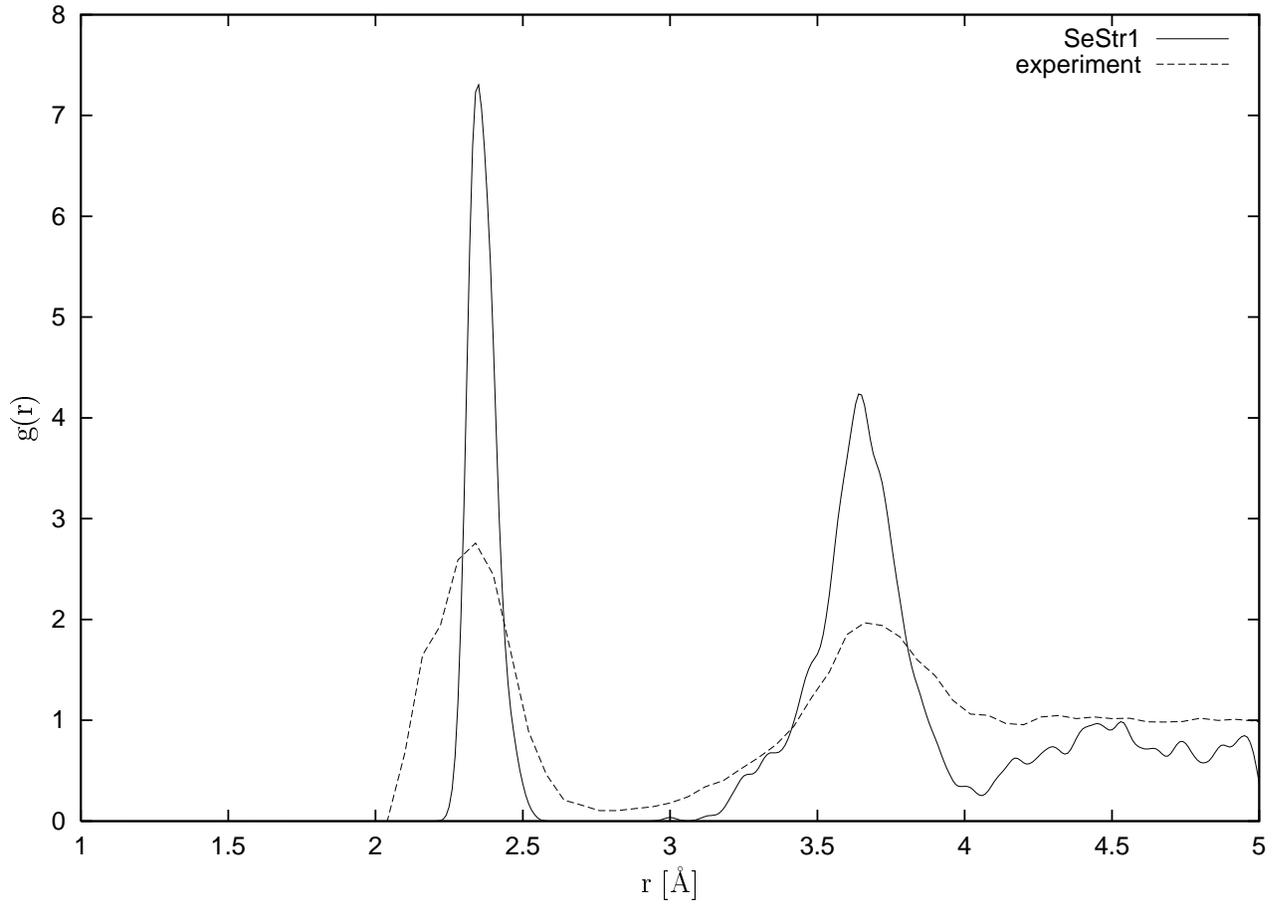}
}
\end{center}
\label{fig2_Se}
\end{figure}

 There are several different crystalline forms of selenium. Basically,
 they consist of chains and eight-membered rings. Typical bond lengths
 are around  2.35~\AA \  while most  of the bond  angle values  can be
 found around $103^{\circ}$.  Snapshot of  
 the amorphous SeStr1 network (bottom half part)
 is shown in Fig.\ 1. We 
   considered 2.8~\AA \ as  an  upper  limit  of bond  length.   Substrates
 remained  similar  to  the  crystal lattice  arrangement  during  the
 bombardment and the relaxation procedure.  The average bond length in
 our a-Se models is $ 2.37 \pm 0.004 $ \AA.  A detailed analysis shows
 that the average  distance between twofold coordinated first-neighbor
 selenium atoms (Se$^2$-Se$^2$) is equal  to 2.35 \AA, \ while in case
 of Se$^2$-Se$^3$  and Se$^3$-Se$^3$ those  values are 2.41 \AA  \ and
 2.47 \AA,  \  respectively (see  Table \ref{TableI_Se}). 
 In order to ignore the
effect of the rough surface on the top of the grown film we identified
two   different    cells:   bulk   and   total    sample   (see   Ref. 5).
  The top  side of the bulk was by  5~\AA \ below the
atom  having  the  largest  $z$  coordinate, furthermore,  bulk does 
not include substrate atoms at the bottom. In
 Fig.\ \ref{fig2_Se}  pair correlation  functions of our  SeStr1 model
 and  an  unconstrained RMC  simulation \cite{jovari03} based on  
experimental data  are  shown in  the
 interval of 1-5~\AA.   \ All the other models  provide similar radial
 distribution functions.  First and second neighbor peak positions are
 similar to trigonal crystalline case but broadened because of torsion
 inside the  chains.  There is a characteristic  inter-chain distance in
 $\alpha$  crystalline  phase  at   3.43~\AA  \  which  is  completely
 disappeared from pair correlation function.

 A histogram  of  calculated  bond  angles  in our model is displayed  in  the  Fig.\
 \ref{fig2_Se}.  The main contribution  to the bond angle distribution
 arises from  angles between  95$^\circ$ and 110$^\circ$.  In $\alpha$
 selenium the bond  angle is $103.1^{\circ}$ which is  larger than the
 average value in our simulations ($102.1^{\circ}$).
 Considering   the    local
arrangements  we can  not distinguish  between deposited  and quenched
models.
The average coordination number in  a-Se is approximately two as shown
in Table  \ref{TableII_Se}.  There is no fourfold coordinated selenium 
atom but we found threefold coordinated atoms (defects) in every models. 
In quenched sample (SeStrQ) we obtained 8 atomic \% while in the 
deposited samples this ratio is higher.

\subsection{Density}

The structures of  different models consist almost of  the same number
of atoms.   For  realistic  density
calculations  one  should   consider only  bulk  densities.   Table\
\ref{TableI_Se} contains these densities of different models which are
between  3.21~g/cm$^3$ and  4.34~g/cm$^3$.  For  crystalline $\alpha$,
$\beta$ and metallic selenium  the densities are equal to 4.4~g/cm$^3$,
4.35~g/cm$^3$  4.8~g/cm$^3$  which  are  larger  than  the  values  we
obtained for  a-Se, i.e.\  our molecular dynamics  simulation provided
lower dense structures.  In order to investigate the homogeneity 
 we divided the structures
prepared by deposition and by rapid quenching  into 
 $\Delta z$=5 \AA \ thick layers. A significant difference was observed
in the local density fluctuation of two models. In  
Fig. \ \ref{fig4_Se} five layer densities of both models are displayed
in function of time. One can conclude that sample prepared by rapid
quenching is more homogen than the deposited counterpart.
This is an  observable difference we obtained for two different 
preparation techniques.

\section{Summary}

  We have developed a molecular dynamics computer code to simulate the
preparation procedure  of a-Se  networks, which are  grown by  a vapor
deposition technique and prepared by  rapid cooling in order to make a
comparison between the atom-by-atom  deposition on a substrate and the
melt-quenching preparation techniques.   The most important difference
we have found between the models prepared at various conditions, is in
local density.   Bond length and bond angle distributions
are very similar in both cases.

\section{Acknowledgments} 

This work has been supported by the Fund OTKA (Grant No.T038191,
T043231)  and
by Hungarian-British  intergovernmental S and T
Programme   (No.  GB-17/2003).   The  computer
simulations  were  partly  carried   out  at  the  Tokyo  Polytechnics
University (Japan). We are indebted to Prof. T. Aoki  (TPU) for providing
us this possibility.



\begin{figure}
\caption{
 Histogram of calculated  bond angles.  The main contribution
to the  bond angle distribution arises from  angles between 95$^\circ$
and    110$^\circ$.   The    average   value    in    our   simulation is
$102.1^{\circ}$. 
}
\vskip 3truecm
\begin{center}
\rotatebox{-90}{
\includegraphics[width=10cm,keepaspectratio]{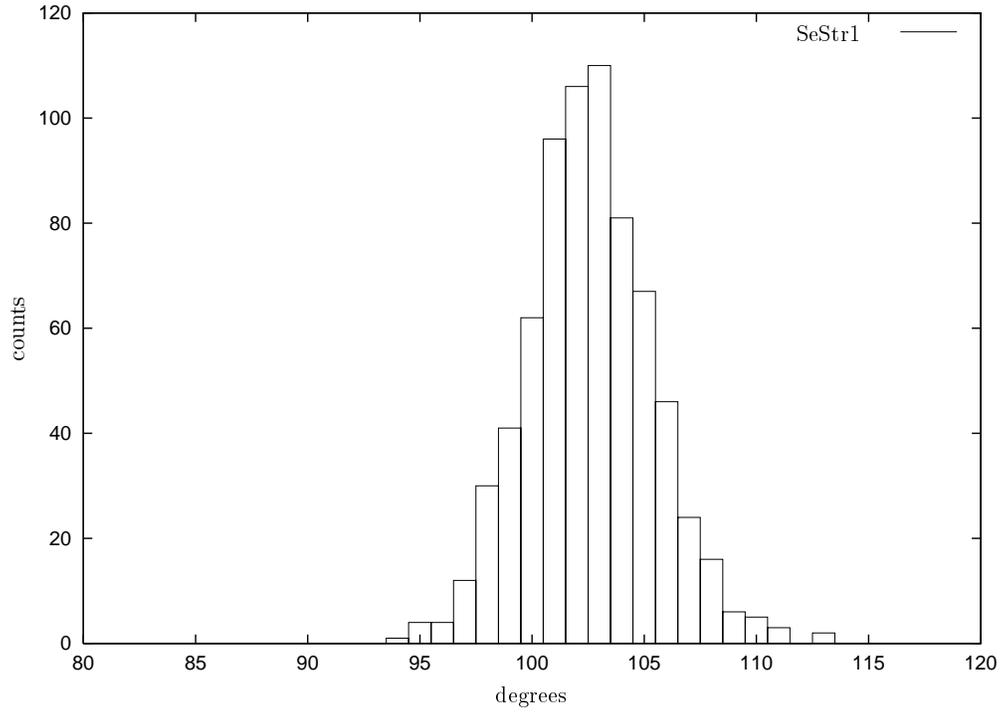}
}
\end{center}
\label{fig3_Se}
\end{figure}

\begin{figure}
\caption{
Density developments of five $\Delta z$=5 \AA \ thick layers
prepared by rapid quenching (top panel) and deposition (bottom panel).  
 A significant difference was observed
in the local density fluctuations of the two models. 
 }
\vskip 3truecm
\begin{center}
\includegraphics[width=12cm,keepaspectratio]{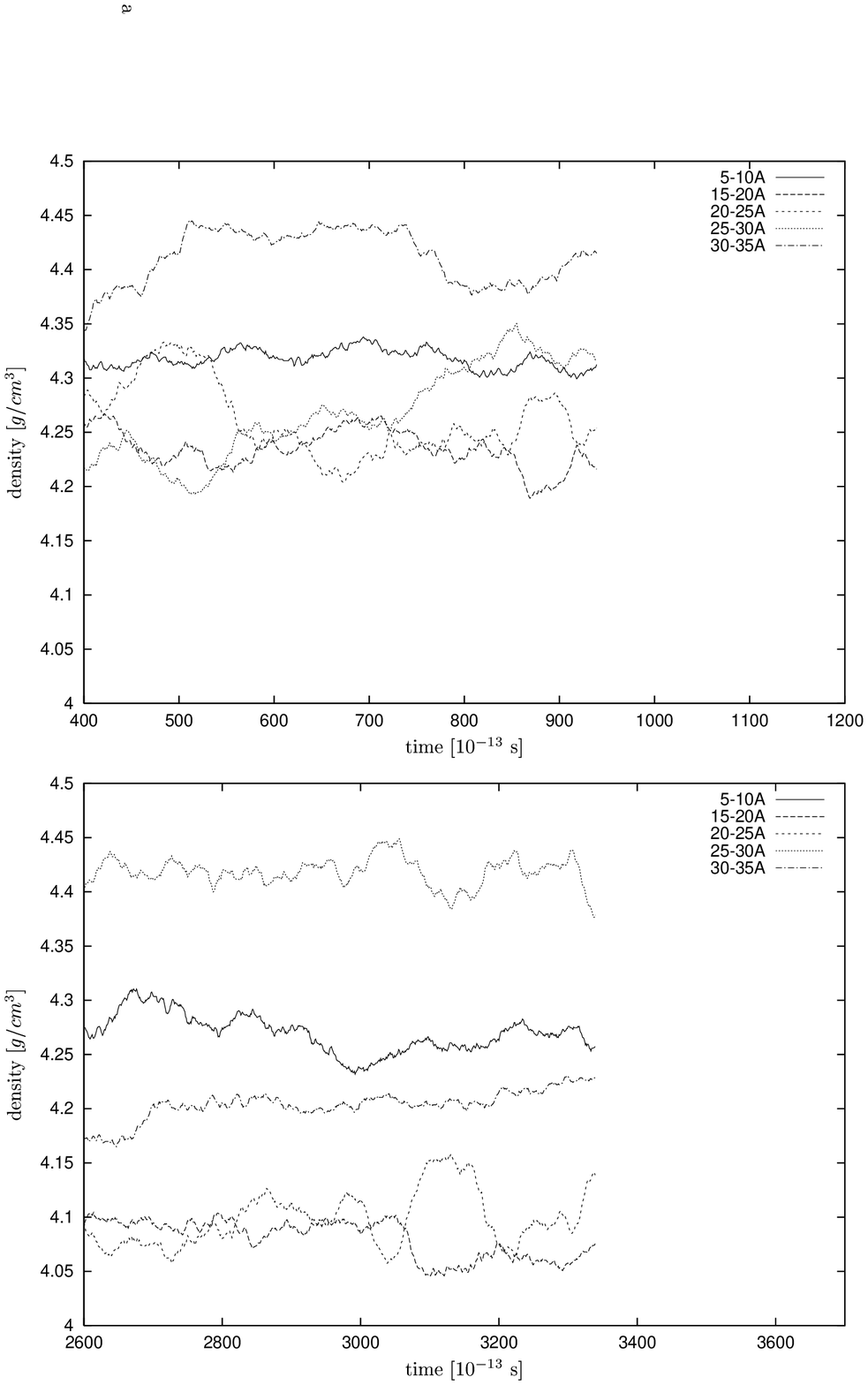}
\end{center}
\label{fig4_Se}
\end{figure}


\begin{table}
\caption{Average and detailed 
 bond lengths  (in \AA) and densities (in g/cm$^3$) 
of different models.
}
\begin{ruledtabular}
\begin{tabular}{|*{8}{r|}}
\hline  Name & Bulk atoms & Total atoms
 & $Se-Se$  & $Se^{3}-Se^{2}$  & $Se^{3}-Se^{3}$  & $Se^{2}-Se^{2}$  & density \\ \hline
SeStr0.1 & 509 & 954  & 2.37  & 2.41  & 2.48  & 2.35  & 3.21 \\ \hline
SeStr1 & 584 & 1016  & 2.37  & 2.41  & 2.47  & 2.35  & 3.73 \\ \hline
SeStr10 & 373 & 822  & 2.37  & 2.41  & 2.46  & 2.35  & 4.34 \\ \hline
SeStrQ & 676 & 1118  & 2.37  & 2.41  & 2.49  & 2.35  & 3.95 \\ \hline
\end{tabular}
\end{ruledtabular}
\label{TableI_Se}
\end{table}

\begin{table}
\caption{
Bond angles (in degrees) and coordination numbers calculated inside bulks.
}
\begin{ruledtabular}
\begin{tabular}{|*{8}{r|}}
\hline  Name
 & $Se-Se^{3}-Se$  & $Se-Se^{2}-Se$  & $Se-Se-Se$ & Z=1 & Z=2 & Z=3 & Z=4 \\ \hline
SeStr0.1 &  100.9  & 102.69  & 102.08  & 2  & 432  & 75  & 0 \\ \hline
SeStr1 &  100.97  & 102.68  & 102.2  & 1  & 516  & 67  & 0 \\ \hline
SeStr10 &  101.08  & 102.68  & 102.25  & 1  & 332  & 40  & 0 \\ \hline
SeStrQ &  100.93  & 102.39  & 102.09  & 0  & 622  & 54  & 0 \\ \hline
\end{tabular}
\end{ruledtabular}
\label{TableII_Se}
\end{table}

\end{document}